\documentclass[useAMS,usenatbib]{mn2e}

\usepackage{epsfig,amsmath,natbib}

\usepackage{aas_macros}
\usepackage{amssymb}
 \usepackage{amsmath}
 \usepackage{dsfont}
 \usepackage{hyperref}
 \usepackage[dvips]{color}

\def\be{\begin{equation}} 
\def\ee{\end{equation}}

\def\mpc{\,{\rm {Mpc}}}

\def\HI{\hbox{H~$\scriptstyle\rm I\ $}} 
\def\H2{\hbox{H$_2$}}

\def\gsim{\lower.5ex\hbox{\gtsima}} 
\def\lsim{\lower.5ex\hbox{\ltsima}} \def\gtsima{$\; \buildrel > \over 
\sim \;$} \def\ltsima{$\; \buildrel < \over \sim \;$} \def\prosima{$\; 
\buildrel \propto \over \sim \;$} \def\gsim{\lower.5ex\hbox{\gtsima}} 
\def\lsim{\lower.5ex\hbox{\ltsima}} 
\def\simgt{\lower.5ex\hbox{\gtsima}} 
\def\simlt{\lower.5ex\hbox{\ltsima}} 
\def\simpr{\lower.5ex\hbox{\prosima}}

\def\gtsima{$\; \buildrel > \over \sim \;$} 
\def\ltsima{$\; \buildrel < \over \sim \;$} 
\def\gsim{\lower.5ex\hbox{\gtsima}} 
\def\lsim{\lower.5ex\hbox{\ltsima}} 
\def\simgt{\lower.5ex\hbox{\gtsima}} 
\def\simlt{\lower.5ex\hbox{\ltsima}} 
\def\simpr{\lower.5ex\hbox{\prosima}}

\def\E3{{\cal E}_{\rm g}^{III}}

\begin{document}

\title{Supermassive black hole ancestors}

\author[A. Petri, A. Ferrara, R. Salvaterra]
{
A. Petri$^{1}\thanks{Email: andrea.petri@sns.it (AP);}$,
A. Ferrara$^{1}\thanks{Email: andrea.ferrara@sns.it (AF);},$
R. Salvaterra$^{2}\thanks{Email: ruben@lambrate.inaf.it (RS);}$
\\
$^{1}$ Scuola Normale Superiore, Piazza dei Cavalieri 7, 56126 Pisa, Italy\\
$^{2}$ INAF, IASF Milano, via E. Bassini 15, I-20133 Milano, Italy
}

\date{13 February 2012}

\pagerange{\pageref{firstpage}--\pageref{lastpage}} \pubyear{2012}

\maketitle
\label{firstpage}

\begin{abstract}
In the attempt to alleviate the difficulties created by their early formation, we study a model in which supermassive black holes (SMBHs) 
can grow by the combined action of gas accretion on heavy seeds and mergers of both heavy ($m_{s}^{h}=10^5 M_\odot$) and light ($m_{s}^{\ell} = 10^2 M_\odot$) seeds. The former result from the direct collapse of gas in $T_{s}^{h} \geq1.5\times10^4$\,K, H$_2$-free halos; the latter are the endproduct of a standard H$_2$-based star formation process. The H$_2$-free condition is attained by exposing halos to a strong ($J_{21} \simgt 10^3$)  Lyman-Werner UV background produced by both accreting BHs and stars, thus establishing a self-regulated growth regime. We find that this condition is met already at $z\sim 18$ in the highly biased regions in which quasars are born. The key parameter allowing the formation of SMBHs by $z=6-7$ is the fraction of halos that can form heavy seeds: the minimum requirement is that $f_{heavy}\simgt 0.001$; SMBH as large as $2\times 10^{10} M_\odot$ can be obtained when $f_{heavy}$ approaches unity. Independently of $f_{heavy}$, the model produces a high-$z$ stellar bulge-black hole mass relation which is steeper than the local one, implying that SMBHs formed before their bulge was in place. The formation of heavy seeds, allowed by the Lyman-Werner radiative feedback in the quasar-forming environment, is crucial to achieve a fast growth of the SMBH by merger events in the early phases of its evolution, i.e. $z \simgt 7$. The UV photon production is largely dominated by stars in galaxies, i.e. black hole accretion radiation is sub-dominant.  Interestingly, we find that the final mass of light BHs and of the SMBH in the quasar is roughly equal by $z=6$; by the same time only 19\% of the initial baryon content has been converted into stars. The SMBH growth is dominated at all epochs $z > 7.2$ by mergers (exceeding accretion by a factor $2-50$); at later times accretion becomes by far the most important growth channel. We finally discuss possible shortcomings of the model.
\end{abstract}

\section{Introduction}
Observations of bright quasars at high redshift $z>6$ (\citealt{Fan1}; \citealt{Fan2}; \citealt{Fan06}; \citealt{Willott09}; \citealt{Willott10}; \citealt{MortlockWarren} and references therein) pose serious challenges to our understanding of their central engine.
In fact, their large bolometric luminosities, $\approx 10^{46} $erg s$^{-1}$, and hard emission spectra are inconsistent with a stellar nature of their energy source. Hence, the current paradigm assumes that the central engine is powered by baryonic accretion onto a central supermassive black hole (SMBH).
However, even this proposal is not immune from additional questions, the most urgent of which concerns the growth time of the SMBH. Assuming that a fraction $(1-\epsilon)$ of the matter is accreted at the Eddington rate, the growth rate of the SMBH can be written as 
\begin{equation}
\frac{d\ln m}{dt} = \frac{1-\epsilon}{\epsilon}\frac{1}{t_E}  
\end{equation}
where $t_E = 4\pi G\mu m_p/\sigma_e c = 0.45$ Gyr is the usual Eddington time. In order to achieve a SMBH of mass $m(t)$ at a cosmic time $t(z)$ corresponding to redshift $z$ it is then necessary to start from a BH seed of mass
\begin{equation}
m_0 = m(t) \exp\left[{-\frac{1-\epsilon}{\epsilon}\frac{t(z)}{t_E}}\right].  
\end{equation}
From the above expression it is clear that assembling the SMBH mass ($m=2\times 10^9 M_\odot$) recently deduced for the most distant quasar ULAS J1120+0641 at $z=7.085$ \citep{MortlockWarren} when $t(z) = 0.77$ Gyr, requires $\ln m_0/M_\odot > 21.4- 1.71(1-\epsilon)/\epsilon$.
For the usually assumed value of $\epsilon = 0.1$, this translates into $m_0 > 400 M_\odot$. Such value is uncomfortably large when compared to the most recent estimates of the mass of first stars, which tend to converge towards values well below $100 M_\odot$ \citep{Greif11, Omukai11, BY11}. Some authors \citep{WillottAJ10} investigated the possibility that even these small seeds ($\lesssim 100M_\odot$) can produce large SMBH masses ($\sim 10^{11}M_\odot$), provided they undergo several major mergers and spend some time accreting at super Eddington rates; current observations support the fact that occasional events of super Eddington growth can occur at $z\sim 6$, and hence, under those conditions, low mass seeds do not pose a serious threat to explain such large SMBH masses.
Alternatively, one has to assume that SMBH growth proceeds through the merging of heavier seeds; this possibility is explored in the present work. The main questions then shift to (i) the formation process of such intermediate mass black holes, and (ii) their number density evolution through time.   

Recent theoretical developments in the study of gas collapse inside dark matter halos propose new mechanisms in which heavy black hole seeds of $\approx 10^5 M_\odot$ can form rapidly  ($\Delta t \approx $ 1\,Myr), and hence lend support the alternative scenario above. In brief, gas inside halos with virial temperature $T_{vir} \simgt 10^4$ K cools almost isothermally provided that it remains molecule-free (in practice, H$_2$-free). \cite{OhHaiman}, and \cite{LodatoNat} have convincingly shown that if H$_2$ formation is inhibited, a primordial gas disk is stable to fragmentation and a single massive object is formed. Hence, a key point for the mechanism to work is that the collapsing halo is exposed to a sufficiently high Lyman-Werner ($10.2  < h\nu/\mathrm{eV} < 13.6)$ soft UV intensity to photo-dissociate the H$_2$ (or the catalyzer H$^-$) via the two-step Solomon process. However, while the UV radiation field favors the formation of heavy BH seeds, it inhibits for the same physical process the formation of stars in minihalos, i.e. halos with viral temperature $< 10^4$ K, which rely on the presence of molecular hydrogen to enable star formation.   

The UV background is produced by both massive stars populating early galaxies and miniquasars\footnote{We generically refer with this term to lower mass, higher redshift counterparts of known SMBHs}. As the Lyman-Werner radiation intensity rises following the formation of first stars and black holes, it is expected that star formation is suppressed in minihalos and continues only in $T_{vir} >10^4$ K halos not fulfilling the requirements for the formation of heavy BH seeds. This process comes to an end when heavy element enrichment of the halos makes gas fragmentation unavoidable \citep{Schneider02, Omukai05, Dopke11}. The interplay between UV background effects and the outcome of the collapse of the earliest structures has been discussed extensively in the literature \citep{Ricotti02,KuhlenMadau05,Abel07,Abel09,Yoshida07,Jeon11} and can hardly be overlooked. The approach in these papers is complementary to ours. In fact, these studies, mostly for reasons related to the limitations set by the limited dynamical range of
numerical simulations, have looked in detail at the feedback effects due to the presence of a miniquasar onto the star formation in the host (mini-)galaxy. Our approach instead, being based on a statistical large-scale scheme (the merger tree), allow us to study the growth of the SMBH in the presence of the collective UV radiation field produced by the galaxies in the QSO environment.  Its novelty consists in the fact that a fully self-consistent approach to this problem implementing the key processes discussed above has not been attempted so far.  

In this paper we intend to specifically address the question of the relative abundances of stellar, light black hole seeds and the heavier ones originating from head-start direct gas collapse including the radiative feedback effects produced by the UV background. Following the canonical procedure in the field \citep{Volonteri1, Volonteri2,TanakaHaiman,Nata11}, we will attack the problem through a Monte Carlo technique based on the extended Press-Schechter {\it Ansatz} \citep{Cole} to simulate the mass distribution and merger history of dark matter halos at various redshifts in a statistical way.  We set the initial conditions for baryons at redshift $z=20$ and track their evolution to $z=6$ coupling analytical prescriptions to the dark matter merger tree. 

We calibrate our models to reproduce the observed properties of $z\approx 6$ quasars in terms of their final SMBH mass and we compare them with the empirical local stellar bulge - black hole mass relation $m_{bulge}/m_{BH}\simeq 10^3$ (\citealt{MH03, HR04}; see also \citealt{ValianteSchneider} for a thorough discussion). We next predict the evolution of the two populations of seed black holes.

Throughout we work in a flat cold dark matter model with a cosmological 
constant ($\Lambda$CDM) cosmology with the cosmological parameters given by the current WMAP7 
\citep{Larson10} best-fit values: $\Omega_{m}=\Omega_{\rm DM}+\Omega_{b}=0.27$, $\Omega_{\Lambda}=1-\Omega_{m}$, $\Omega_{b}h^2=0.02249$, $h = 0.704$, $\sigma_8 = 0.8$  and $dn_s/d\ln k =0$.

\section{Model}
\label{model}
In this Section we present the basic features of our model. The evolution and dynamics of the dark matter field is treated in a statistical way by the merger history of a certain dark matter halo of mass $M$ which collapsed at redshift $z$. Each merger event between DM halos is characterized by the dynamical friction timescale, $t_{merge}$, that depends, to a first approximation, only on the ratio $x=M_>/M_<$ of the colliding halo masses. To compute $t_{merge}$ we use Chandrasekhar theory \citep{Mo} which states that 
\begin{equation}
\label{chandra}
H(z)t_{merge}=0.234\frac{\eta x}{\ln{(1+x^2)}},
\end{equation}
where $\eta$ is a circularity parameter encoding the eccentricity of the orbit decay. We define an event with $t_{merge}<\langle\Delta t\rangle$ ($t_{merge}>\langle\Delta t\rangle$), with $\langle\Delta t\rangle$ being the mean time resolution of the computation discussed below, as a major (minor) merger event. If the gas contained in the dark matter halo potential wells can cool efficiently (i.e. the timescale of the relevant cooling processes $t_{cool}$ is smaller than the Hubble time $t_H$), stars and black holes of various masses can form; the dark matter merger dynamics subsequently drives the assembly of progressively heavier objects. 
\subsection{Merger tree}
The merger tree is a computational realization of the extended Press-Schechter (P-S) {\it Ansatz}, which gives an approximate expression for the mass distribution function of collapsed dark matter objects in the universe (see \citealt{LaceyCole93} for a more detailed treatment). Take a halo of mass $M_1$ at redshift $z_1$: according to P-S, the mass distribution of its progenitors at some higher redshift $z_2>z_1$ is given by the following expression 
\begin{equation}
\label{extps}
\frac{dN}{dM_2}(z_2 \rightarrow z_1)=\frac{M_1}{M_2^2}f_{PS}(\nu_{12})\left\vert\frac{d\ln \nu_{12}}{d\ln{M_2}}\right\vert
\end{equation}
where 
\begin{equation}
\label{nu}
\nu_{12}= \frac{\delta_c(z_2)-\delta_c(z_1)}{\sqrt{S(M_2)-S(M_1)}} 
\end{equation}
and $S(M)$ is the variance of the cosmic density field smoothed on a mass scale $M$ with a sharp $k$ filter. Assuming a spherical collapse model we have
\begin{equation}
\label{pressfct}
f_{PS}(\nu)=\sqrt{\frac{2}{\pi}}\nu\exp{\left(-\frac{\nu^2}{2}\right)}
\end{equation}
In order to obtain the mass distribution $dN/dM_2$ from a numerical algorithm,  one needs to specify a mass resolution $M_{res}$; to keep computational times reasonable, all details below $M_{res}$ are ignored. It is useful to define two additional quantities, that are the mean number of progenitors $P$ in the mass interval $M_{res}<M_2<M_1/2$, defined as
\begin{equation}
P=\int_{M_{res}}^{M_1/2}\frac{dN}{dM_2}dM_2,
\end{equation}
and the fraction of mass $F$ of the final object in progenitors below the resolution limit, defined as 
\begin{equation}
F=\int_0^{M_{res}}\frac{dN}{dM_2}\frac{M_2}{M_1}dM_2
\end{equation}
To split a halo of a certain mass and redshift in its progenitors we follow \cite{Cole}.  Such procedure starts from the parent halo $(M_1,z_1)$; we then pick a redshift step $\Delta z_1$ such that $P\ll 1$ in order to ensure that the halo has at most two progenitors (binary mergers) at redshift $z_1+\Delta z_1$. Given a random number $R$ in the interval $0<R<1$, if $R>P$ the halo is not split and its mass is reduced to $M_1(1-F)$ (masses below $M_{res}$ are considered as accreted gas from the surroundings during $\Delta z_1$). If instead $R\leq P$ the halo is split in two and a random value $M_2$ in the range $M_{res}<M_2<M_1/2$ is generated from the distribution (\ref{extps}); the two progenitors are assigned masses $M_2$ and $M_1(1-F)-M_2$. 

Albeit this procedure seems to work at a first sight, there is a shortcoming: by definition, when one draws randomly a mass $M_2<M_1/2$, the mass of the other fragment is automatically chosen as $M_1(1-F)-M_2$, i.e. the two values are equally probable, in contrast with the probability distribution predicted by eq. (\ref{extps}). Moreover, such algorithm uses equation (\ref{extps}) only for $M_2<M_1/2$ and ignores its predictions for $M_2\geq M_1/2$: this results in a halo bias towards the low mass end. The fact that this procedure tends to underestimate the mass of the most massive progenitors is a well known problem; to overcome this problem \cite{Cole} suggest the following modification of the distribution function: 
\begin{equation}
\label{extpscole}
\frac{dN}{dM_2} \rightarrow \frac{dN}{dM_2}G\left(\frac{\sigma_2}{\sigma_1},\frac{\delta_1}{\sigma_1}\right)
\end{equation}
The dependence on the first argument allows the distribution of fragments to be modified, while the dependence on $\delta_1/\sigma_1$ allows the splitting rate to change; only modifications at first order in $\ln{G}$ are retained yielding 
\begin{equation}
G\left(\frac{\sigma_2}{\sigma_1},\frac{\delta_1}{\sigma_1}\right)=G_0\left(\frac{\sigma_2}{\sigma_1}\right)^{\gamma_1}\left(\frac{\delta_1}{\sigma_1}\right)^{\gamma_2}
\end{equation}
The values $G_0, \gamma_1,\gamma_2$ were chosen so to match the code output to that of the Millennium Simulation\footnote{\url{http://www.mpa-garching.mpg.de/millennium}}, which is one among the largest cosmological simulations available to date (see also \citealt{Springel05}). According to this analysis we set $G_0=0.57, \gamma_1=0.38, \gamma_2=-0.01$. 

Beside the above biasing problem, merger trees have another important limitation: matter elements with $m<M_{res}$ are actually accreted by the merged halos but are ignored by the numerical procedure. These unresolved mass elements can account for a non-negligible fraction of the total dark matter mass ($\sim 90\%$ at $z=20$) of the simulation. We claim that all this missing mass can be reintegrated in the tree without any consequences for structure formation, since this mass is split in very small elements of mass smaller than $M_{res}$. The reason is that star and BH formation can occur only in halos  above a certain mass threshold (discussed in the remainder of this Section) corresponding to $M\gtrsim 10^5 M_\odot > M_{res} $ at $z=20$ and increasing at lower $z$. Hence, unresolved structures only bring in dark matter and gas, but no stars or BHs.

We now turn to the discussion on how we populate individual halos of the tree with black holes and stars and we follow their joint evolution. Before we do this though we briefly divert to discuss some important features of the mechanism leading to the formation of the heavy BH seeds.

\subsection{Heavy black hole seeds}
Several authors \citep{BrommLoeb,LoebRasio,OhHaiman,LodatoNat,VolonteriRees,Regan09} have proposed that heavy ($\sim 10^5 M_\odot$) black hole seeds can be formed from direct gas collapse, perhaps passing through a very short intermediate stellar-like phase. 

However, hydro-simulations by \cite{BrommLoeb}, suggest that heavy seeds resulting from the direct collapse of gas clouds can be formed only if the UV background in the Lyman-Werner bands is high enough to photodissociate molecular hydrogen. In fact, if $H_2$ is present, the gas cools rapidly, thus strongly decreasing the gas fragmentation mass scale. Such simulations showed that a mass of $\sim 10^6 M_\odot$ can condense within a radius ($\sim 1$\,pc) where they had to stop due to resolution limits; such radius is unfortunately still much larger than the Schwarzschild radius of the system. \cite{LoebRasio} arrived at similar conclusions estimating the minimum mass the central collapsed object should have in order to stabilize the surrounding disk against fragmentation. 

More recently, \cite{OhHaiman} and \cite{LodatoNat}, suggested that the formation of a heavy black hole seed proceeds via the gravitational collapse of accretion disks in H$_2$-free halos and investigated the conditions under which these disks can sink into the center of the potential well via the Toomre instability. We stress that the formation of a disk is necessary in order for gravitational collapse to occur, because the typical adiabatic index of the gas, calculated e.g. by \cite{SpaansSilk} is too low for spherical collapse to proceed without fragmentation. \cite{VolonteriRees} tried to overcome the numerical limitations preventing the study of the collapse up to the Schwarzschild radius through semi-analytical methods. They were then able to investigate the small scale physics at the end stage of the collapse. The most remarkable result they obtain is that in these systems, because of the high mass of the collapsed object ($\gg M_\odot$), by the time nuclear reactions are able to proceed gravity is too strong in order for the collapse to be halted. When the temperature reaches extreme values ($\sim 10^9$\,K) weak processes with the emission of neutrinos occur and this causes catastrophic cooling that leads to the formation of a black hole. Hence, it appears that - broadly speaking - the formation of heavy seeds from gas collapse is physically possible and it is regulated by the strength of the UV background governing the cooling ability of the gas. Although this statement is hardly conclusive and additional study is required to put it on a more firm ground, contingent evidence encourages us to adopt it in the following analysis. We check that the gas mass available in the halo exceeds the heavy BH seed mass $m_s^h = 10^5 M_\odot$.

\begin{figure}
\begin{center}
\includegraphics[width=1\columnwidth]{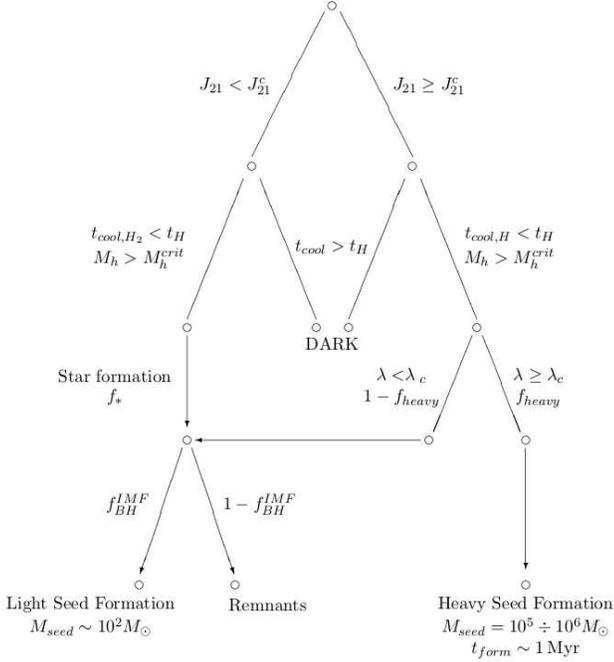}
\caption{Schematic representation of the conditional physical processes considered}
\label{seedscheme}
\end{center}
\end{figure}

\subsection{Seeding the tree}
We assign to each halo of mass $M$ on the highest level of the tree ($z=20$) a baryonic mass
\begin{equation}
M_b=\frac{\Omega_b}{\Omega_m}M.
\end{equation} 
We also assign this baryonic mass to all the halos in the lower levels that have no parent, that is to say all those halos that are too small to undergo further splitting. After setting this, we need to decide in which of these halos the gas can cool efficiently and hence stars or black holes can be formed. Gas gravitational collapse inside the host halos can occur only if the cooling time is shorter than the Hubble time ($t_{cool}<t_H$). Radiative cooling is provided by two distinct channels: (i) atomic \HI lines, or (ii) H$_2$ molecule rotational bands; whether one or the other cooling channel is enabled, depends on intensity of radiative background in the Lyman-Werner (LW) band ($\langle h\nu \rangle\sim 12.42$\,eV). Using the standard normalization $J_\nu=J_{21} \times 10^{-21}$erg s$^{-1}$cm$^{-2}$Hz$^{-1}$\,sr$^{-1}$, one can define a critical value, $J_{21}^c$, separating the two weak and strong field regimes. These are discussed separately in the next; the seeding prescriptions are graphically summarized in Fig. \ref{seedscheme}.

\subsubsection{Weak UV field limit}
If $J_{21}\lesssim J_{21}^c$ cooling can proceed only in halos whose mass exceeds a threshold value $M_{crit}$, given by 
\begin{equation}
M_{crit}(z)= 1.5 \times 10^6 h^{-1} M_\odot \left[\left(\frac{T_{s}^\ell}{1200\mathrm{K}}\right)\left(\frac{10}{1+z}\right)\right]^{3/2}.
\end{equation}
\cite{TanakaHaiman} suggest to use the value $T_{s}^\ell = 1200$\,K-Physically, the excitation temperature of the rotational levels of $H_2$, the main cooling agent at zero metallicity, would give $T_s^\ell \gtrsim 600$\,K; however a more detailed calculation (e.g. see \citealt{Mo})  yields $T_s^\ell \gtrsim 1200$\,K. The critical mass is essentially set by the ability of minihalos to cool via the self-synthesized H$_2$. In this case, star formation proceeds with efficiency $f_*$. It is yet unclear if a single massive ($\simgt 100 M_\odot$) star forms or rather the final configuration is a normal stellar population. Although uncertainties remain (see Introduction), a common opinion is emerging, supported by several lines of observational evidence, that the first stars were not as massive as previously thought. We take a conservative approach assuming that a stellar cluster of total mass $m_{star}=f_* m_{b}$ is formed according to a Salpeter Initial Mass Function (IMF)  
\begin{equation} 
\label{salp}
\phi(m)\propto m^{-\alpha},
\end{equation}
with $\alpha=2.35$. 
Stars leaving behind BH remnants are those in the mass range $40M_\odot < m < 140M_\odot$ and $m>260M_\odot$ (in the intermediate range the star ends its life as a pair-instability supernova). The lower cutoff $m_{cut}$ for the fragments mass fixes the normalization constant and so we can integrate eq. (\ref{salp}) to obtain a value for $f_{BH}^{IMF}$ as follows
\begin{equation}
f_{BH}^{IMF}(m_{cut})=\int_{BH}\phi(m) dm 
\end{equation}
Integration yields the following values: $f_{BH}^{IMF}(20M_\odot)=0.35$ and $f_{BH}^{IMF}(1M_\odot)=0.006$; the cut at $20 M_\odot$ is suitable for collapsing gas inside metal-free primordial halos;  $m_{cut} =1M_\odot$ applies to halos which have been polluted by metals from previous generations and have a smaller fragmentation scale. For simplicity, we assume that every halo which has hosted stars at the previous (i.e. higher redshift) merger tree step is metal polluted and therefore $f_{BH}^{IMF}=0.006$  

\subsubsection{Strong UV field limit}
If $J_{21}\gtrsim J_{21}^c$, molecular hydrogen is photodissociated and cooling can proceed via atomic H lines (and metal lines, if present). If cooling is effective (i.e. a condition requiring a halo virial temperature $T_{s}^{h} \geq1.5\times10^4$\,K) then one must evaluate the halo spin parameter, $\lambda$, to infer the fate of the gas. Because the situation is still not well understood from the theoretical point of view, we make the following simplifying assumption: if $\lambda > \lambda_c$ then the gas settles into an accretion disk that fuels the formation of a heavy black hole seed of mass $m_{heavy}$ via Toomre instability, on a short timescale, typically  $1$\,Myr. If instead $\lambda < \lambda_c$ then the disk does not form, the gas cloud collapses nearly spherically and fragments, again forming stars with efficiency $f_*$. We recall that this scenario is mainly supported by the evidence that, given the typical values of the adiabatic index calculated by \cite{SpaansSilk}, a spherical gas cloud is very likely to fragment; thus,  the formation of a disk is needed to allow the formation of a massive central object. We can translate the spin threshold criterion into the fraction of halos that can form heavy seeds, $f_{heavy}$, via the log-normal probability distribution $p(\lambda)$ \citep{WarrenQuinn}: 
\begin{equation}
f_{heavy}(\lambda_c)=\int_{\lambda_c}^{\infty}p(\lambda) d\lambda
\end{equation}  
The value of $\lambda_c$ depends on small scale details (see \citealt{Regan09} for a thorough discussion) such as the gas temperature, $T_{gas}$, the collapsed baryon fraction, $f_d$,  and metallicity, which are all rather uncertain; we then keep $\lambda_c$ (or alternatively $f_{heavy}$) as an additional free parameter. We stress that the heavy seeds are thought to be formed on a very short timescale $t_{form}\sim 1$\,Myr compared to the mean time step used $\langle\Delta t\rangle \sim t_H/15$ with $t_H$ ranging from 1.4\,Gyr at $z=6$ to 0.3\,Gyr at $z=20$. The details of the formation hence hopefully should not affect our results.

\subsection{Ultraviolet background}
\begin{figure}
\begin{center}
\includegraphics[width=1\columnwidth]{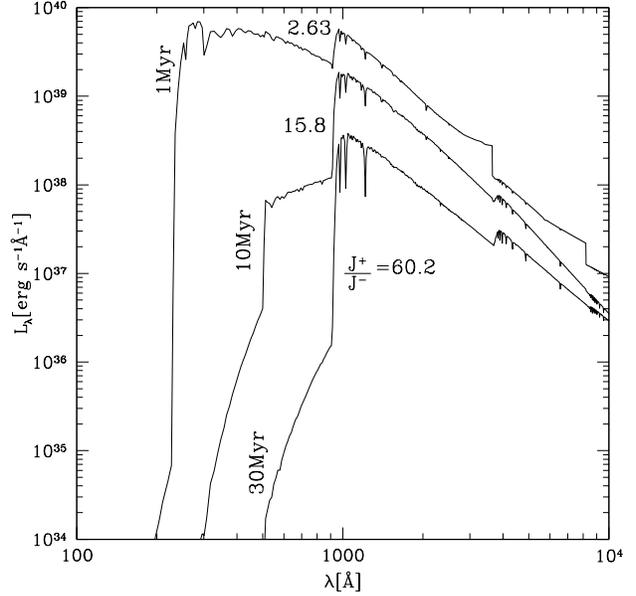}
\caption{Stellar continuum spectrum for instantaneous star formation mode, normalized to a star cluster of mass $m=10^6 M_\odot$ with metallicity $Z=0.001$, and an IMF$\propto m^{-2.35}$ with $m_{cut}=1 M_\odot$. Labels shown the value of the spectral jump, $J^+/J^-$ below and above 1 Ryd.}
\label{stars}
\end{center}
\end{figure}

How strong could be the LW background in the biased region in which SMBHs we observe at $z \approx 6$ are built? A minimal and yet robust estimate can be derived as follows.
There is now strong observational evidence that at $z=6$ cosmic reionization was essentially complete. Thus, one can compute directly the mean intensity of the {\it ionizing} (cosmological) UV background, $J_\nu^+$, required to keep the intergalactic hydrogen ionized by requiring that at least a photon per baryon is produced, $n_\gamma/n_b=1$. However, this estimate represents a lower limit to $J_{\nu}^+$ as it does not account for recombinations occurring in overdense regions. Recent numerical works \citep{Mitr11} suggests to take $n_\gamma/n_b=10$. It is then easy to show that 
\begin{equation}
4\pi J_{\nu}^+(z) =  \frac{h_P c  \langle\rho_H(z)\rangle}{m_p} \left(\frac{n_\gamma}{n_b}\right)
\end{equation}  
or  
\begin{equation}
J_{\nu}^+(z) =  0.14 (\Omega_b h^2)(1+z)^3 \left(\frac{n_\gamma}{n_b}\right),
\end{equation}  
which yields $J_{21}^+(z=6) = 0.97(n_\gamma/n_b) \approx 10$. In the previous equations,
$h_P$ is the Planck constant, $c$ is the speed of light, $m_p$ is the proton mass, and 
$\langle\rho_H(z)\rangle$ is the mean density of hydrogen contributing 76\% of the total gas mass. However the intensity of LW radiation, which is not absorbed by hydrogen atoms, can be larger by a factor equal to the inverse of the escape fraction of ionizing photons $f_{esc}^{-1} \approx 10^{1-2}$ multiplied by the jump across 1 Ryd in the intrinsic spectrum of a typical stellar population (see Fig. \ref{stars}), which is a function of stellar age\footnote{Results obtained with STARBURST99 \citep{Leitherer99,VL05}, software available at \\ \url{http://www.stsci.edu/science/starburst99/}} for the IMF used here. As we see from the Figure, the jump is in the range $2.6-60.2$ for a young stellar population and increases further with age. We can safely conclude that close to reionization, we expect average LW intensities as $> J_{21} \approx 10^3$. We however expect the (local) values of $J_{21}$ be higher in the star forming (and high quasar activity) regions we consider being higher than this average value, and a more detailed calculation is needed to correctly estimate the local LW background. These details are explained in the subsequent paragraphs 2.4.1 and 2.4.2. 

In practice, though, one needs to properly follow the UV radiation sources during their evolution to obtain a more precise estimate.   
We model the emission of sources in the merger tree through a redshift-dependent specific physical emissivity $\epsilon(\nu,z)=L_{\nu}(z)(1+z)^3/V_{box}$, where $V_{box}=M_h/\bar{\rho}_m$ is the comoving volume corresponding to a linear perturbation of mass $M_h$ that we use to determine the simulated volume, and $L_\nu =\sum_i L_{\nu,i}$ is the total luminosity produced by the sources in the LW band. For the $M_h = 10^{12} M_\odot$ halo we are simulating the comoving volume is $V_{box}= 29 \mpc^3$.

Lyman-Werner photons interact with hydrogen molecules present in the intergalactic medium and are absorbed. Assuming a \H2 relic abundance $\approx 10^{-5}$ one typically gets \citep{CiardiFerrara} an optical depth\footnote{During the evolution, as $J_{21}$ rises, most of the \H2 will be cleared by photodissociation with a consequent decrease of the opacity; we neglect this complication.}  $\tau_{H_2}\approx 3$.  We assume that inside $V_{box}$ the radiation field is homogeneous. This is justified by the fact that the mean free path of LW photons is much larger than the physical size of the simulated volume. The background can be computed by properly accounting from redshifted radiation from higher redshift sources, via the standard formula:
\begin{equation}
\label{ciardi}
J_{\nu}(z)=\frac{c}{4\pi}\int_z^{z_{on}}\epsilon\left(\nu\frac{1+z'}{1+z},z'\right) e^{-\tau_{H_2}}\left(\frac{1+z}{1+z'}\right)^3 \left\vert\frac{dt'}{dz'}\right\vert dz'
\end{equation}
where $z_{on}$ is the source turn-on redshift. 

Apart from more exotic and rather unconstrained radiation sources as dark matter particle annihilations/decays \citep{Mapelli06,Valdes10}, there are two types of sources that contribute to the UV radiation background, i.e. stars and accreting black holes.

\subsubsection{Stars}
UV radiation is copiously produced by massive stars, particularly if young and metal-poor.
For example a metal free population of stars with a $1 M_\odot - 500 M_\odot$ Salpeter IMF, produces $10^{4.469}$ \H2 photodissociating photons per baryon into stars while on the ZAMS 
(compared to $10^{4.355}$ in H-ionizing photons). Given that the narrow width of the LW band (2.4 eV), the stellar spectrum is assumed to be flat within it; its time evolution  (Fig. \ref{stars}) has been derived through the public stellar population synthesis code  {\tt STARBURST99} for an instantaneous\footnote{This implies that stars form on a timescale $t_{form}\ll\langle\Delta t\rangle$.} burst of star formation with metallicity $Z=10^{-3} \approx 0.05 Z_\odot$ for the Salpeter IMF with the two different low mass cut-offs $m_{cut}=(1,20) M_\odot$ introduced above.   

\subsubsection{Accreting black holes}
In comparison to stars, black hole emission is currently less understood and we need to make three simplifying assumptions in order to compute it. These are: 
(i) each heavy black hole that is surrounded by a cloud of gas accretes mass at the Eddington rate
\begin{equation}
\dot{m}_{BH}=f_{duty}\left(\frac{1-\epsilon}{\epsilon}\right)\left(\frac{m_{BH}}{t_{E}}\right)
\end{equation}
where $t_{E}=0.45$\,Gyr; the radiation conversion efficiency is taken to be $\epsilon=0.1$ and we introduced a duty cycle factor $f_{duty}=0.5-1$ to account for the fact that gas accretion may proceed at a lower rate \citep{Marconi04,Brusa09,Fiore12}. We note that the parameter $f_{duty}$ contains a degeneracy between the quasar lifetime and the accretion rate. For example, two accretion events with $f_{duty}=0.5$, one proceeding at half the Eddington rate for time interval $\Delta t$, and the other proceeding at the Eddington rate for half of the time interval $\Delta t$ will result in the same net accreted mass. When the black hole has a mass $m_{BH}$ we assume that its luminosity coincides with the Eddington limit $L_{E}(m_{BH})=1.3\times 10^{40}\,\mathrm{erg}\, \mathrm{s}^{-1} (m_{BH}/100M_\odot)$; (ii) light black holes do not accrete gas and hence do not contribute to luminosity. This is supported by the results by \cite{Alvarez08}) who have showed that this type of BHs are characterized by very low ($\approx 10 
^{-12} M_\odot$ yr$^{-1}$) accretion rate partly due to radiative feedback and also because they spend most of their lifetime in low-density regions. This setting is different than that of \cite{TanakaHaiman}, where they include the possibility that light seeds do accrete, principally due to the fact that these seeds form in the center of their host halo. In our setting however, light seeds are scattered in the stellar bulge, where the gas low density prevents them from accreting significantly. We shold also note that \cite{Milosavljevic09} arrived at similar results as \cite{Alvarez08}:
both simulations used the Bondi-Hoyle-Lyttleton formula to estimate the accretion rate. \cite{Milosavljevic09} also suggest that even feedback-suppressed systems may undergo short episodes of potentially super-Eddington accretion, but in this paper we neglect this possibility; (iii) All accreting black holes are characterized by a non-thermal $\propto 1/\nu$ emission spectrum in the band 10\,eV$<h\nu<$10\,keV; this sets the normalization
\begin{equation}
L_{\nu}(m_{BH})=\frac{1}{3\ln{10}}\frac{L_{E}(m_{BH})}{\nu}
\end{equation}
The above spectra for the two types of sources are then used to calculate the value of $J_{21}$ through eq.(\ref{ciardi}). 
Finally we need to fix the chosen flux threshold value, $J_{21}^c$, for H$_2$ photodissociation. This value depends on the spectrum shape and we refer to recent work to obtain a numerical estimate. \cite{BrommLoeb} find $J_{21}^c\approx 10^5$ for a quasar power-law spectrum and $J_{21}^c\lesssim 10^3$ for a $10^4$\,K thermal spectrum that approximates well a stellar-dominated background. More recently, the quasar threshold has been decreased by an order of magnitude \citep{Omukai}; such finding has been confirmed and refined by a more careful estimate of the LW opacity by \cite{WolcottGreen}. These authors suggest to use the threshold values $J_{21}^c=10^3$ for a stellar background and $J_{21}^c=4.3 \times 10^3$ for a power-law quasar background. In our simulation, the actual UV background intensity and shape represents a combination of the stellar and accreting black hole contributions and therefore in such a mixed situation it is not obvious which of the two thresholds applies. Lacking a more definite insight, we adopted the prescription that H$_2$ photodissociation occurs when one of the two thresholds is reached, independently of spectral shape of the dominant component. This choice is not crucial in practice, as we will see later on, BH radiation is always largely sub-dominant with respect to the stellar one. To summarize, the numerical parameters we use are ($J_{21}^c, T_{s}^{\ell}, T_{s}^{h},  m_{s}^{\ell}, m_{s}^{h}, f_*$) = ($1-4.3 \times 10^3, 1200\,\mathrm{K},  1.5 \times10^4\,\mathrm{K}, 10^2 M_\odot, 10^5M_\odot, 0.1$).

\section{Simulations}
\label{simulations}
\begin{figure}
\begin{center}
\includegraphics[width=1\columnwidth]{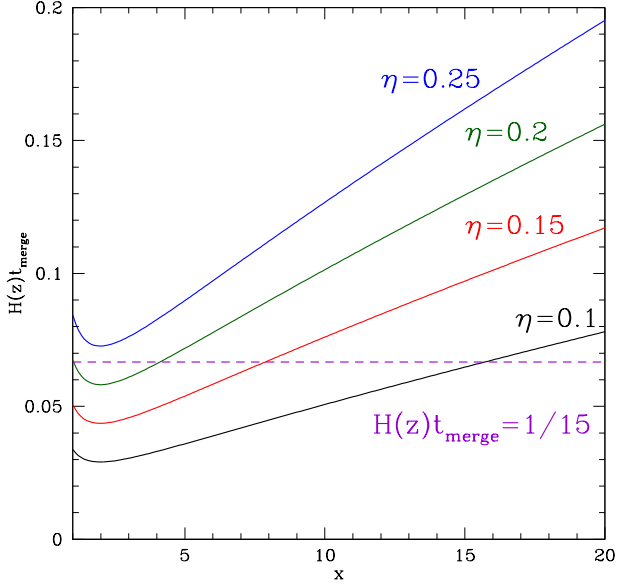}
\caption{Merging timescale in units of the Hubble time (\ref{chandra}) as a function of the mass ratio $x=M_>/M_<$ for different values of the circularity parameter $\eta$. Also shown is the value $H(z)t_{merge}=1/15$ corresponding to our mean timestep in a tree with $z\in[6,20]$.}
\label{merge}
\end{center}
\end{figure}

In this Section we will briefly describe the simulation implementation. The merger tree is built very similarly to \cite{Cole} and it 
determines the dark matter halo mass function evolution with redshift. We follow the tree in the redshift interval $z\in[6,20]$ with a mass resolution of $M_{res}=10^5M_\odot$; the adopted redshift step is $\Delta z=0.7$, corresponding to a mean timestep $\langle \Delta t\rangle=40$\,Myr (or $\langle H\Delta t\rangle=1/15$). This choice is motivated by two reasons: the first is that, since the heavy black hole seeds are thought to form on a timescale $t_{form}=1-10$\,Myr, we can ignore the uncertain formation process details. The second is that the timestep should be larger than the timescale for a major merger, which cannot be arbitrary small; taking a mean circularity parameter $\eta=0.2$, from Fig. \ref{merge} we identify $x=4$ as the critical mass ratio that discriminates between major and minor mergers. Motivated by the recent discovery of \cite{MortlockWarren} we included also some runs with 20 timesteps in $z\in[7.08,20]$, a mean time resolution $\langle \Delta t\rangle=30$\,Myr and a critical mass ratio $x=3$. With such high mass and time resolution we can computationally afford to compute the merger history of a large final SMBH host halo of $M_h=10^{12}M_\odot$ at $z=6$. 
The choice of $z=20$ as the initial redshift is dictated by computational reasons: with this choice, we achieve a mass resolution of $10^5M_\odot$, essentially equal to the critical mass (at $z=20$) discriminating between star-forming and ''dark'' halos, in which the gas is not able to cool. Pushing the initial redshift to $z=30-40$, would require a further increase the resolution to retain the ability to discriminate star forming halos. We do not expect our assumption of $z=20$ as the initial condition to affect greatly our results because the timesteps become very small at high redshift, and hence the BH seeds do not accrete a significant amount of mass before that time.

We track the evolution of baryons in each halo in the following different components: (i) a central BH, built from heavy seeds merging and gas accretion (see below; we neglect direct accretion of stars by the BH); (ii) a bulge component, containing stars, light BHs and gas available for BH accretion or further star formation, (iii) the satellites, i.e. sub-halos resulting from minor mergers, which contain heavy and light seeds, stars and gas. Some discussion on the merger prescription is in order at this point. During a {\it major} 
merger all the contents of the light halo sink into the center of the heavy halo. In particular, heavy holes merge, whereas light holes, stars and gas are simply added to the bulge; the contents of the heavy halo satellites remain the same. For {\it minor} mergers instead,  the heavy halo remains unchanged, whereas the contents of the light halo are added to the heavy halo as satellites. When two black holes merge, we always assume that the merging timescale is controlled by the dark matter dynamical friction timescale, i.e. black holes merge only during major mergers. Throughout this work we neglect the final stage of the merger which is regulated by the gravitational radiation loss timescale. The latter is much smaller than all other relevant times scales for heavy seeds mergers, but it is unclear whether the same condition  holds also for light - heavy seeds mergers, due to the final parsec barrier. Neglecting this not yet well understood process (see for example \citealt{Nixon11, Preto11}) introduces some uncertainties in our model, but its influence on our results, whose uncertainty is predominantly of statistical nature, is likely to be subdominant.
In all the cases considered only the heavy BHs in the bulges are allowed to accrete at the Eddington rate (with duty cycle $f_{duty}$) until they run out of fuel; on the contrary, no accretion takes place in the periphery, and hence BHs in satellites do not emit. Multiple merger events in the tree have been modelled as a sequence of binary mergers occurring in a random order.  

Finally a mechanism that can cause the merging of light and heavy holes has been included: the light holes in the bulge experience dynamical friction due to stars and can sink onto the central BH (if present). The calculations for this sinking timescale, along with the definition of major and minor mergers, are analogous to those for the dark matter merging timescale $t_{merge}$ so we do not repeat them here. Using these prescriptions we set the initial conditions for baryons at $z=20$ and track their evolution to $z=6$ (or $z=7.08$) to verify if a central black hole mass of $\sim 10^9 M_\odot$ can be put together.

\section{Results}
\label{results}
\begin{figure}
\begin{center}
\includegraphics[width=1\columnwidth]{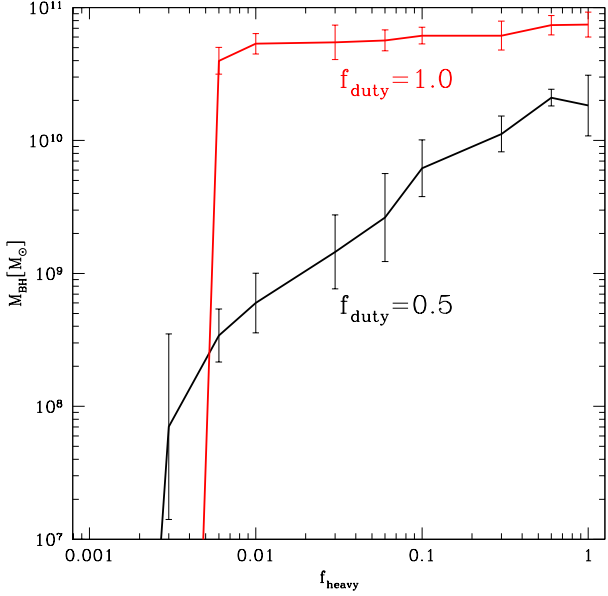}
\caption{Final black hole mass at $z_f=6$ as a function of the heavy seeding fraction, $f_{heavy}$, for two values of the duty cycle, $f_{duty}$. Error bars show 1$\sigma$ dispersion around the mean over 100 merger tree realizations.}
\label{spin}
\end{center}
\end{figure}

\begin{figure}
\begin{center}
\includegraphics[width=1\columnwidth]{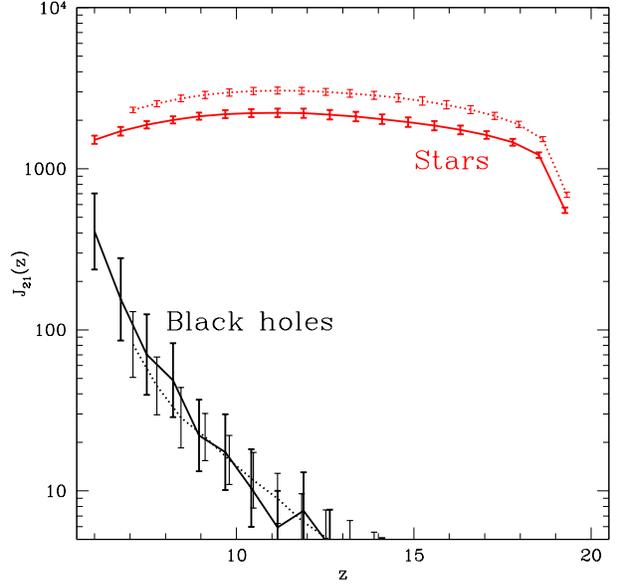}
\caption{Evolution of the UV Lyman-Werner radiation field $J_{21}(z)$. Plotted are the stellar (red lines) and black hole (black lines) contributions for $z_f=6$ (solid) and $z_f=7.08$ (dotted) for the case $M_h=10^{12}M_\odot ,f_{duty}=0.5$, $f_{heavy}=0.01$. Error bars show 1$\sigma$ dispersion around the mean over 100 merger tree realizations.}
\label{back}
\end{center}
\end{figure}

In this Section we show the results obtained from the model described above. Predictions concern mostly the case for a SMBH hosted by either a $M_h=10^{12}M_\odot$ or $M_h=10^{13}M_\odot$ halo at $z_f=6$. However, motivated by the recent discovery of the most distant quasar \citep{MortlockWarren}, we have also performed some runs ending at $z_f=7.08$.  
The first point we want to make is that the chosen values of the accretion duty cycle $f_{duty}$ and of the critical spin parameter $\lambda_c$ (or equivalently, in our model, of the heavy seeding fraction $f_{heavy}$) set the order of magnitude of the final black hole mass obtained at $z_f=6$, as can be seen in Figure \ref{spin}. 
If accretion occurs at the Eddington rate ($f_{duty}=1$), the black hole mass is almost independent of $f_{heavy}$ and, beyond the sharp cutoff at $f_{heavy}\sim 0.01$, it sets to a value in the range $M_{BH} = 6-7\times 10^{10}M_\odot$. Thus, extremely large black holes, exceeding the current estimates, can be formed under these conditions. The more realistic case, $f_{duty}=0.5$, yields a final central BH mass at $z_f=6$ in the $10^{8-10}M_\odot$ range, strongly dependent on the value of $f_{heavy}$. Table \ref{outline} contains a summary of the final black hole and bulge masses obtained for various $f_{duty}$ and $z_f$, for a fixed value of $f_{heavy}=0.01$. 
\begin{table}
\begin{center}
\begin{tabular}{|c|c|c|c|}
\hline\hline
$f_{duty}$ & $\log{(M_{BH}/M_\odot)}$&$\log{(M_{bulge}/M_\odot)}$ \\ \hline \hline
\multicolumn{3}{|c|}{$M_h=10^{12}M_\odot,\, M_{res}=10^5M_\odot$} \\ \hline
\multicolumn{3}{|c|}{$z_f=6$} \\ \hline
$0.5$  & $8.5 \pm 0.2$ &$11.02 \pm 0.08$  \\ 
\hline 
$1.0$ & $10.6 \pm 0.1$ &$10.7 \pm 0.2$ \\ \hline 
\multicolumn{3}{|c|}{$z_f=7.08$} \\ \hline
$0.5$ & $7.8 \pm 0.3$ &$11.00 \pm 0.08$ \\ 
\hline 
$1.0$ & $10.0 \pm 0.4$ &$10.8 \pm 0.1$ \\ \hline \hline
\multicolumn{3}{|c|}{$z_f=6, M_h=10^{13}M_\odot,\, M_{res}=2 \times 10^6M_\odot$} \\ \hline
0.5 & $9.1 \pm 0.3$ & $12.06 \pm 0.06$\\ \hline
1.0 & $11.6 \pm 0.1$ & $11.8 \pm 0.09$ \\ \hline
\end{tabular}
\caption{Outline of the results for $f_{heavy}=0.006$; the final black hole mass $M_{BH}$ and the mass of the surrounding baryonic bulge $M_{bulge}$ are displayed. There is also a run with $M_h=10^{13}M_\odot$. Error bars show 1$\sigma$ dispersion around the mean over 100 merger tree realizations.}
\label{outline}
\end{center}
\end{table}
Let us now examine in more detail the preferred model in which $f_{duty}=0.5$ and $f_{heavy}=0.01$ by considering the predicted evolution of the radiation background, the star/BH formation rates, and the masses of the various baryonic components. 
Looking at Figure \ref{back}, we see that the Lyman-Werner radiation field, and hence H$_2$ photodissociation, is dominated by the stellar background.  The specific intensity, $J_{21}(z)$, rises very rapidly and up-crosses the critical threshold $J_{21}^c =10^3$ already at $z\approx 18$, thus triggering the formation of heavy black hole seeds. This is particularly important, as it shows that in the highly biased regions in which quasars are born, the UV radiation from galaxies is sufficient to allow the formation of the heavy seed required to build them. In this model the black hole UV background component does never become supercritical. The fact that an early epoch of star formation is necessary for the formation of heavy seeds was also suggested by \cite{BrommLoeb}. We note that even if the average cosmological background reaches the critical threshhold value $\sim 10^3$ only at the epoch of reionization, the local one in star forming regions is sigificantly bigger, and reaches the threshold for heavy seed formation much earlier.

\begin{figure}
\begin{center}
\includegraphics[width=1\columnwidth]{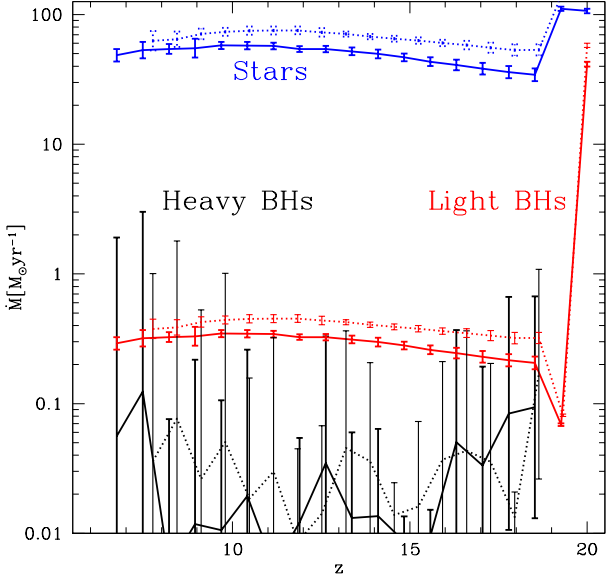}
\caption{Formation rate $\dot{M}$ of stars (blue lines), light (red), and heavy (black) black hole seeds as a function of redshift for $z_f=6$ (solid) and $z_f=7.08$ (dotted) and a model with $M_h=10^{12}M_\odot, f_{duty}=0.5$, $f_{heavy}=0.01$. Error bars show 1$\sigma$ dispersion around the mean over 100 merger tree realizations.}
\label{form}
\end{center}
\end{figure}

The early appearence of heavy seeds can be better appreciated from Fig. \ref{form}, where the various formation rates are compared. The light BH seeds formation rate has an initial peak, it decreases and finally stabilizes to a much lower and relatively constant value, $0.2-0.3 M_\odot$ yr$^{-1}$, as a result of the progressive metal enrichment that forces the light seed formation to proceed with a lower $f_{BH}^{IMF}=0.006$, as discussed in Sec. \ref{model}. As expected the light BH formation rate closely tracks the star formation rate; the heavy BH seeds formation rate instead shows considerable fluctuations around a lower mean value. Once formed the heavy BH seeds grow by accretion and merging with other heavy seeds, according to the merging prescriptions introduced above. The peaks in the stars and light BH's formation rates are a numerical artifact related to the way in which we modeled metallicity. We made the following simplifying assumption: if a halo did not experience star formation before, we consider it as metal-free and we set $20M_\odot$ as a IMF lower cut. If instead star formation already occurred, then we assume that its metallicity is above the critical one and cut the IMF at  $1M_\odot$. This is admittedly the simplest possible hypothesis and this leads to the peaks in Fig. 6. The correct solution is probably smoother but this effect is only a marginal one affecting the very phases of the system evolution.  

\begin{figure}
\begin{center}
\includegraphics[width=1\columnwidth]{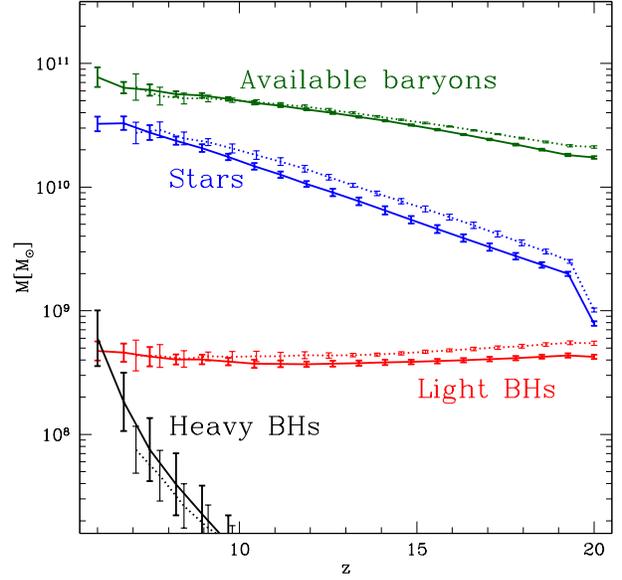}
\caption{Redshift evolution of the stellar, black holes, and available baryons for $z_f=6$ (solid) and $z_f=7.08$ (dotted), for the model with $M_h=10^{12}M_\odot, f_{duty}=0.5, f_{heavy}=0.01$. Error bars show 1$\sigma$ dispersion around the mean over 100 merger tree realizations.}
\label{mass}
\end{center}
\end{figure}

The final mass of the two compact populations, light BHs and the central SMBH, is roughly equal by $z=6$, as seen from Fig. \ref{mass} where we present also the evolution of the redshift evolution of the baryonic components. However, a nice feature of the results is that both populations are subdominant with respect to stars by $2-3$ orders of magnitude. As we will discuss in more detail in a moment,  this is broadly consistent with observation of the stellar bulge - black hole mass relation for high-$z$ quasars. Baryons are not completely converted into stars even though star formation has proceeded at very sustained rates ($> 50 M_\odot$yr$^{-1}$) for essentially the entire Hubble time; we find that 19\% of the initial baryon content has been converted into stars. The baryonic content is continuously replenished by infalling gas  (in our scheme, baryons contained in very small, starless halos with mass below the resolution mass, $M_{res}$). 

\begin{figure}
\begin{center}
\includegraphics[width=1\columnwidth]{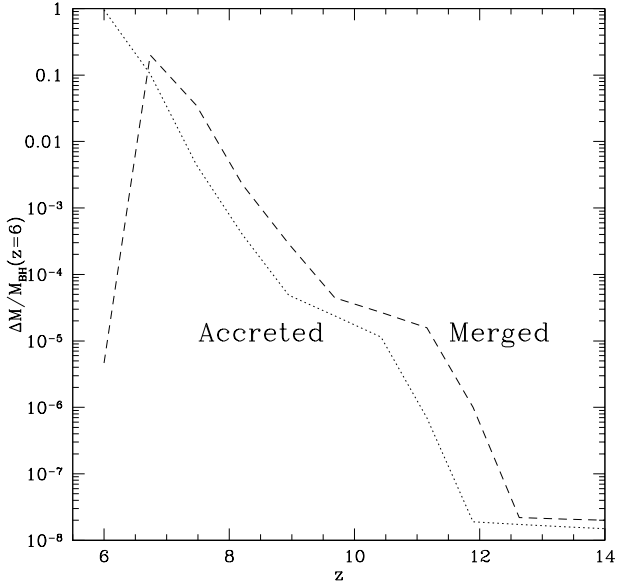}
\caption{Fraction of the final SMBH mass, $\Delta M_{acc}/M_{BH}$, gained by either accretion or merging in each tree redshift step for the model with $z_f=6$, $M_h=10^{12}M_\odot, f_{duty}=0.5, f_{heavy}=0.01$.}
\label{versus}
\end{center}
\end{figure}

\begin{figure}
\begin{center}
\includegraphics[width=1\columnwidth]{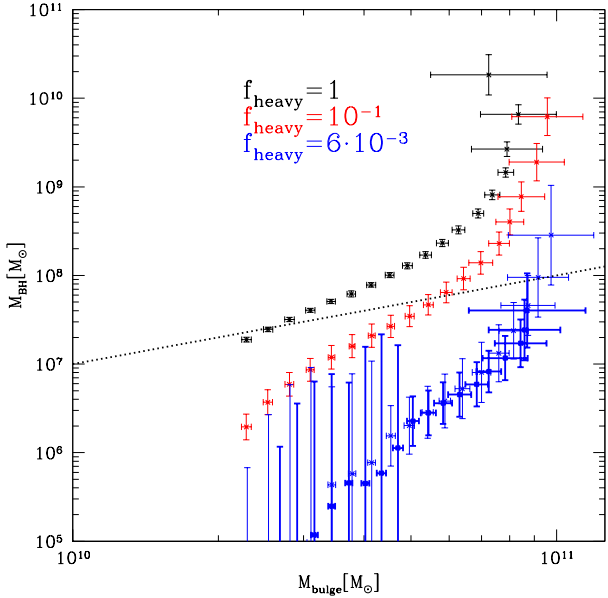}
\caption{$M_{BH}-M_{bulge}$ relation for a tree realizations with $z_f=6$ (thin points) at each tree level for three different values of $f_{heavy}$, as indicated. For the best fit value $f_{heavy}=6\times 10^{-3}$ we show also the case with $z_f=7.08$ (thick points). The dotted line is the local empirical relation $M_{BH}=0.001M_{bulge}$.}
\label{starbh}
\end{center}
\end{figure}

The SMBH mass $M_{BH}$ at $z=6$ is set by the combination and balance between gas accretion (we do not consider accretion of stars) and BH mergers along the tree hierarchy. A natural question is then which of these two process prevails along the build-up history. This question is answered by Fig. \ref{versus} where we plot the fraction of the final SMBH mass, $\Delta M_{acc}/M_{BH}$, gained by either accretion or merging in each tree redshift step. These curves has been obtained by tracking the evolution of the central BH mass along the most massive progenitor branch of the merger tree and by defining the "merged" mass as $\Delta M_{mer}=M_{BH}(z)-\Delta M_{acc}$. For the accreted mass in a timestep $\Delta t$ (or in a redshift step $\Delta z$) we used the usual expression
\begin{equation}
\frac{\Delta M_{acc}(\Delta z)}{M_{BH}(z)}=e^{f_{duty }\frac{1-\epsilon}{\epsilon}\frac{\Delta t(z)}{t_E}}-1
\end{equation}
The SMBH growth is dominated at all epochs $z\ge 7.2$ by mergers, which dominate the fractional mass gain by a factor $2-50$ depending on redshift. Between $z=7.2$ and $z=6$, the relation is reversed and accretion becomes by far the most important
growth channel, with mergers progressively becoming negligible with time.  During this late evolutionary stage the final black hole mass grows by a factor of nearly three.  As already shown, these results depend on the unknown  fraction of halos that can form heavy seeds, $f_{heavy}$.  If one poses that the local stellar bulge - black hole mass relation holds also at the high redshifts of concern here, we can take advantage of the empirical observational constraint $M_{BH}/M_{bulge}\sim 10^{-3}$ (see \citealt{ValianteSchneider}) to obtain a more realistic value for $f_{heavy}$.  Decreasing $f_{heavy}$ values tend to improve the agreement between the model results at $z=6$ and the local empirical relation (Fig. \ref {starbh}): the best fit is obtained for $(f_{duty} ,f_{heavy}) = (0.5, 0.006)$. Independently on $f_{heavy}$, though the relation is steeper than linear, implying that SMBHs formed before their bulge was in place. This trend is in line with the results of recent numerical simulations investigating the nature of quasar hosts at $z=6$ \citep{Khandai11}.
%
 
\section{Conclusions}
\label{discussion}

In the attempt to alleviate the difficulties created by the observed rapid formation of SMBHs, which are already in place only 770 Myr after the Big Bang, we have proposed  a model in which the hole can grow by a combined action of gas accretion and mergers of both heavy ($m_{s}^{h}=10^5 M_\odot$) and light ($m_{s}^{\ell} = 10^2 M_\odot$) seeds. We stress that, in our picture, the heavy seeds are formed in the center of their host halos and hence are able to accrete gas, whereas the light seeds are scattered in low density regions and hence do not accrete gas efficiently. The former result from the direct collapse of gas in $T_{s}^{h} \geq1.5\times10^4$\,K, H$_2$-free halos; the latter are the endproduct of a standard H$_2$-based star formation process which also depends on the enrichment conditions of the gas. The molecular-free condition is attained by exposing halos to a strong ($J_{21} \simgt 10^3$)  Lyman-Werner UV background produced by both accreting BHs and, predominantly as we have shown, by stars. Quite noticeably, UV radiation in the biased region in which quasars live can in principle establish a self-regulated joint evolution (schematically represented in Fig. 1) of star formation and black hole formation/growth:  fluxes exceeding the
critical value $J_{21}^c$ favor the formation of heavy BH seeds, but they simultaneously depress star formation in small halos, and hence their UV photon production.  With our model, based on a well-tested merger tree scheme and a treatment of the Lyman-Werner field intensity produced by both stars in galaxies and black holes, we have been able to follow all these  processes in detail.     

We have shown that the key parameter allowing the formation of SMBHs by $z=6-7$ is the fraction of halos that can form heavy seeds: the minimum requirement is that, $f_{heavy}\simgt 0.01$; SMBH as large as $2\times 10^{10} M_\odot$ can be obtained when $f_{heavy}$ approaches unity. The precise value of this parameter is very hard to derive from first principles \citep{Regan09}, as it depends on unknown environmental properties (gas temperature and metallicity, collapsed baryon fraction). However, the mere existence of SMBHs places the above solid lower limit to this quantity. Independently on the value of $f_{heavy}$, though, we find that at high-$z$ the  stellar bulge - black hole mass relation is steeper than the local one, implying that SMBHs formed before their bulge was in place. The formation of these heavy seeds is then crucial to achieve a fast growth of the SMBH by frequent merger events in the early phases of their evolution, $z \simgt 7$. Their formation is allowed by the fact that the Lyman-Werner radiation field intensity rises very rapidly and overshoots the critical threshold $J_{21}^c$ in the quasar-forming environment already at $z \approx 18$; such photon input is almost completely dominated by stars in galaxies. This UV flux, on the other hand,  considerably quenches star formation in mini-halos, but we find that this effect is negligible as far as the UV intensity is concerned, i.e. weak feedback strength. 

Interestingly, the final mass of light BHs and of the SMBH in the quasar is roughly equal by $z=6$; by the same time only 19\% of the initial baryon content has been converted into stars. The SMBH growth is dominated at all epochs $z > 7.2$ by mergers (exceeding accretion by a factor $2-50$); beyond that epoch accretion becomes by far the most important growth channel.

In spite of the novel results obtained, the present work should be seen only as a useful guideline for future and more 
in-depth work. A number of additional physical effects should be included along with a more detailed description of
many aspects of the problem. 

The treatment of the small scale physics of black hole mergers can be improved in several ways, for example considering the influence of the stellar density profile \citep{Volonteri1} and gravitational kicks effects \citep{TanakaHaiman,Baker} which might cause the ejection of a fraction the BH from the host galaxy. Gas accretion could be modeled in a somewhat more realistic way through the use of  Bondi's formula; however, the validity of such approach is unclear. In addition it would require specific assumptions on the gas density profile.  It would also be interesting to attack some of the unresolved aspects related to disk fragmentation induced by the presence of metals.
Finally, we neglected the radiative feedback of the quasars on the surrounding gas: \cite{ValianteSchneider} showed that by means of this mechanism star formation in the host galaxy can be efficiently quenched and the gas expelled. This process should become important only in the final phases of the evolution, when the accretion rates become extremely large and therefore it should not have dramatic effects on the high-$z$ evolution with which we are mainly concerned here. 

Future observational developments that could in principle put some constraints on our model include the activities of the International X-ray Observatory (IXO) for which we aim at making detailed predictions in the future, and those of the Laser Interferometer Space Antenna (LISA) that should be capable to detect gravitational wave bursts originating from black hole merger events at high redshifts \citep{Sesana11}.

\section*{Acknowledgements}
We thank S. Cole for technical help with the merger tree code, R. Schneider, M. Volonteri and Z. Haiman for useful discussions.
\bibliographystyle{mn2e}
\bibliography{article}
\label{lastpage}
\end{document}